\documentstyle[preprint,aps]{revtex}

\begin{document}
\draft
\title{Implementation for Solving Random Satisfiability Problems through CNOT-based
circuits in a NMR\ Quantum Processor}
\author{Xinhua Peng$^{\thanks{%
E-mail addresses: xhpeng@wipm.ac.cn or xinhuapeng555@hotmail.com;
xwzhu@wipm.ac.cn; klgao@wipm.ac.cn; Fax: 0086-27-87199291.}}$, Xiwen Zhu and
Kelin Gao}
\address{State Key Laboratory of Magnetic Resonance and Atomic and Molecular Physics,%
\\
Wuhan Institute of Physics and Mathematics, The Chinese Academy of Sciences,%
\\
Wuhan, 430071, People's Republic of China}
\maketitle

\begin{abstract}
We give a general method of construting quantum circuit for random {\it %
satisfiability} (SAT) problems with the basic logic gates such as
multi-qubit controlled-NOT and NOT gates. The sizes of these circuits are
almost the same as the sizes of the SAT formulas. Further, a parallelization
scheme is described to solve random SAT problems efficiently through these
quantum circuits in {\it nuclear magnetic resonance} (NMR) ensemble quantum
computing. This scheme exploits truly mixed states as input states rather
than pseudo-pure states, and combines with the topological nanture of the
NMR spectrum to identify the solutions to SAT problems in a parallel way.
Several typical SAT problems have been experimentally demonstrated by this
scheme with good performances.
\end{abstract}



\section{Introduction}

A SAT problem, associated with the combinatorial search problem\cite{Garey},
is well known as one of the most difficult {\it nondeterministic polynomial}
(NP) problems. As the first problem to be shown NP-complete\cite{Cook}, the
SAT problem is a central problem in combinatorial optimization. On
conventional computers, one expects that no efficient algorithm to solve
random SAT problem in polynomial time can be found, whereas quantum
computers can simultaneously evaluate all search states ({\it quantum
parallelism}) which endows quantum computation with the extraordinary
capabilities superior to its conventional counterpart. As a result, of
pratical interest is whether the computational resources for NP problems can
be reduced on quantum computers. The most striking results so far were
obtained for certain NP and some difficulty problems\cite
{Deutsch,Cleve,Shor,Grover,Hogg}. For instance, Hogg\cite{Hogg} put forward
a highly structured quantum algorithm to solve a 1-SAT and maximally
constrained $k$-SAT problem in a single step with probability 100\% on a
ground state input $\left| 00...0\right\rangle ,$ which have been
experimentally implemented in NMR ensemble quantum computing\cite{Zhu,Peng}.

In standard quantum computation, quantum computers are machines that operate
and control pure quantum states according to quantum mechanics. In turn, the
proposals of NMR ensemble quantum computing\cite{Gershenfeld,Cory1} involve
manipulations of highly mixed states, i.e., pseudo-pure states, and has
achieved rapid devolopment in recent years. At the beginning of the
proposals, it has been proved that an expectation value quantum computer
(EVQC) (e.g., a NMR quantum computer) is capable of solving NP-complete
problems in polynomial time, such as the SAT problems\cite{Cory1}. Such an
EVQC can not only judge if the SAT formula is satisfiable but also count the
number of satisfying assignments. Furthermore, in 1998, Knill et al. \cite
{Knill}proposed the ``one-qubit'' model of quantum computation where the
initial state is a truly mixed state, i.e., the first qubit is 0 and the
other qubits are completely random, and described the power of this model.
In the same year, Madi et al.\cite{Madi} presented a parallelization scheme
for quantum computation with mixed superposition states by using the
operators in spin Liouville space, which has been utilized to solve certain
SAT problems when only the existence of a solution is considered.

In this paper, a general method is given to construct quantum circuits for
random SAT problems with C$^{n}$-NOT and NOT gates. The size of the circuit
is evaluated to be approximately the size of the given SAT formula.
Analogous to the ``one-qubit'' model of quantum computation, we present a
parallelization scheme of solving random SAT problems through these
constructed circuit with a truly mixed state. The information contained in
the line-splittings is used to extract the explicit knowledge of satisfying
assignments from inspection of multiplet of just one qubit. In principle,
hence, a NP-complete SAT problem is solved in polynomial time on a NMR
quantum information processor which invokes {\it quantum parallelism}.
Meanwhile, we have experimentally implemented several typical SAT problems
on a 3- or 4-qubit NMR quantum information processor by this scheme. These
experiments exclude the extreme difficulty of the pseudo-pure state
preparation because truly mixed superposition states for parallelizaion acts
as inputs instead of pure states. In practice, this analogous
parallelization scheme can be applied to other problems of intrest\cite
{Grover,Xiao}.

\section{Scheme}

\subsection{The Satisfiability Problem and Quantum Circuit}

A SAT problem\cite{Garey} is described in terms of a logic formula $F$ in
conjunctive normal form (CNF), consisting of $m$ logical clauses $\left\{
C_{\mu }\right\} _{\mu =1,...,m}$ over a set of $n$ Boolean variables $%
\left\{ x_{i}=0,1\right\} _{i=1,...,n}$ with $0=FALSE$ and $1=TRUE$. Each
clause is the logic OR connection ($\vee $) of some chosen variables or
their negations, e.g., $C_{\mu }=\left( x_{i}\vee \overline{x_{j}}\vee
...\vee x_{k}\right) $. The logic formula $F$ can be expressed as the
logical AND connection ($\wedge $) of all clauses, i.e., 
\begin{equation}
F\left( x_{1},x_{2},...,x_{n}\right) =\wedge _{\mu =1}^{m}C_{\mu }.
\end{equation}
A solution to the SAT problem is an assignment $\left\{ x_{i}\right\}
_{i=1,...,n}$ by specifying a value for each variable $x_{i}$, satisfying
the formula $F$, that is, $F\left( x_{1},x_{2},...,x_{n}\right) =1$. An
important restricted case of SAT is $k$-SAT$\ $when all the clauses have
exactly $k$ variables, which is NP-complete for $k\geq 3$\cite
{Garey,Nielsenbook}. An example of the 3-SAT problem with three variables ($%
n=3$)and three clauses ($m=3$) is the propositional formula $\left(
x_{1}\vee x_{2}\vee x_{3}\right) \wedge (x_{1}\vee x_{2}\vee \overline{x_{3}}%
)\wedge \left( \overline{x_{1}}\vee x_{2}\vee x_{3}\right) $. This problem
has five solutions, e.g., $\left\{ x_{3}=1,x_{2}=1,x_{1}=0\right\} $, an
assignment with the bit representation 110. The study of random $k$-SAT
problems for $k\geq 3$ has played a major role in both classical and quantum
computational sciences.

From the achievements of quantum computing, it is possible to convert the
Boolean operations into a sequence of unitary transformations by designing a
resultant {\it reversible} gate 
\begin{equation}
U_{F-C-NOT}:U_{F-C-NOT}\left| x,x_{0}=0\right\rangle =\left| x,F\left(
x\right) \right\rangle
\end{equation}
where the input $\left| x\right\rangle =\left| x_{n}\right\rangle \left|
x_{n-1}\right\rangle ...\left| x_{1}\right\rangle $ is the control registe
and a work one-bit $I_{0}$ whose initial state $\left| x_{0}\right\rangle
=\left| 0\right\rangle $ is the target qubit to store the computational
output of the formula $F$. Here, variable $x_{i}$ is viewed as the $i$-th
bit $I_{i}$ whose state is represented by $\left| x_{i}\right\rangle $ which
can be equal to either $\left| 0\right\rangle \equiv FALSE$ or $\left|
1\right\rangle \equiv TRUE$. Ref. \cite{Lee} gives a practical method of
constructing quantum Boolean circuits (QBC) for Boolean functions by using
NOT and general multi-bit controlled-NOT gates. The Toffoli gate has been
shown to a well-known {\it reversible} gate sufficient to implement all
Boolean functions\cite{Deutsch,Nielsenbook,Preskill}. For example, the
operation of ($x_{i}\wedge x_{j}$) is performed by a Toffoli gate (see
Figure 1(a)), while an alternative, simple performance of ($x_{i}\wedge
x_{j} $) is to combine a Toffoli gate with NOT gates (see Figure 1(b))
because the NOT gate is itself reversible.

Therefore, the $U_{F-C-NOT}$ gate for the CNF formula $F$ of Eq. (1) can be
easily constructed with QBCs. Due to the OR connection of some variables in
each clause, the unitary transformation to compute every clause is realized
by one generalized multi-bit controlled-NOT gate $CNOT\left( \left\{
I_{i}\right\} ,I_{s_{\mu }}\right) $ sandwiched by NOT gates as 
\begin{eqnarray}
U_{C_{\mu }} &=&\left( \prod_{x_{j}\in \left\{ \overline{x_{j}}|C_{\mu
}\right\} }N_{x_{j}}\right) \left( \prod_{x_{i}\in \left\{ all\text{ }%
x_{i}|C_{\mu }\right\} }N_{x_{i}}\right) CNOT\left( \left\{ I_{i}\right\}
,I_{s_{\mu }}\right)  \nonumber \\
&&\left( \prod_{x_{i}\in \left\{ all\text{ }x_{i}|C_{\mu }\right\}
}N_{x_{i}}\right) \left( \prod_{x_{j}\in \left\{ \overline{x_{j}}|C_{\mu
}\right\} }N_{x_{j}}\right) \left( N_{s_{\mu }}\right) \\
&=&\left( \prod_{x_{k}\in \left\{ x_{k}|C_{\mu }\right\} }N_{x_{k}}\right)
CNOT\left( \left\{ x_{i}\right\} ,I_{s_{\mu }}\right) \left( \prod_{x_{k}\in
\left\{ x_{k}|C_{\mu }\right\} }N_{x_{k}}\right) \left( N_{s_{\mu }}\right)
\end{eqnarray}
where $\left\{ I_{i}\right\} $ is a set of the control bits which includes
the $i$-th bit iff variable $x_{i}$ or $\overline{x_{i}}$ appears in the
clause $C_{\mu }$, $I_{s_{\mu }}$ is the target bit whose state $\left|
s_{\mu }\right\rangle $ is initialized into $\left| 0\right\rangle $ as an
additional scratchpad to store the value of the clause $C_{\mu }$, and $%
N_{x_{i}}$ represents the NOT gate on the {\it i-}th bit. The first pair of $%
\prod_{x_{j}\in \left\{ \overline{x_{j}}|C_{\mu }\right\} }N_{x_{j}}$ on the
bits $I_{j}$ are used to realize the $\overline{x_{j}}$ operations in the
clause $C_{\mu }$ with the second $\prod_{x_{j}\in \left\{ \overline{x_{j}}%
|C_{\mu }\right\} }N_{x_{j}}$ restoring the input to its original value for
the next use. Whereas the second pair of $\prod_{x_{i}\in \left\{ all\text{ }%
x_{i}|C_{\mu }\right\} }N_{x_{i}}$ on the bits $I_{i}$ are the requirement
of the OR operation in the clause in the light of Figure 1 (b). In general, $%
m$ additional qubits $s_{\mu }$ are required. The $U_{F-C-NOT}$ gate to
realize the SAT formula $F$ is obtained by multiplying all clause gate $%
U_{C_{\mu }}$ followed by one $CNOT\left( \left\{ I_{s_{\mu }}\right\}
,I_{0}\right) $ gate to perform the AND operation of all $m$ clauses, i.e.,

\begin{equation}
U_{F-C-NOT}=CNOT\left( \left\{ I_{s_{\mu }}\right\} ,I_{0}\right) \prod_{\mu
=1}^{m}U_{C_{\mu }}
\end{equation}
where $\left\{ I_{s_{\mu }}\right\} $ is a set of all scratchpad bits $%
I_{s_{\mu }}$ storing the results of $m$ clauses, respectively, as the
control bits and the value of the CNF formula $F$ is given through the work
bit $I_{0}$ whose initial state $\left| x_{0}\right\rangle $ is set to $%
\left| x_{0}\right\rangle =\left| 0\right\rangle $. Thus a SAT problem can
be simulated by CNOT-based circuits. For example, the QBC to simulate a
3-SAT mentioned above is shown in Figure 2 (a). Usually, this $U_{F-C-NOT}$
gate in Eq. (4) originally generated by this construction can be further
simplified to get the most effecient circuit by some logic rules\cite
{Lee,Iwama} .

In the circuit model of computation, the number of elementary gates of a
circuit and the number of time steps required are often checked to evaluate
the size of the circuit so as to determine its complexity. We first
concetrate on the $k$-SAT problem because of its importance.

For the simplest case of 1-SAT problems, the gate sequence to simulate a
given 1-SAT formula $F_{1}$ with $m$ clauses can be simplified as 
\begin{equation}
U_{F-C-NOT}=\left( \prod_{x_{j}\in \left\{ \overline{x_{j}}|F_{1}\right\}
}N_{x_{j}}\right) CNOT\left( \left\{ I_{i}\right\} _{\in \left\{ all\text{ }%
x_{i}|F_{1}\right\} },I_{0}\right) \left( \prod_{x_{j}\in \left\{ \overline{%
x_{j}}|F_{1}\right\} }N_{x_{j}}\right) ,
\end{equation}
which requires one C$^{m}$-NOT gate and several NOT gates. Llikewise for $k$%
-SAT\ problems with one clause $C_{1}$, 
\begin{equation}
U_{F-C-NOT}=\left( \prod_{x_{k}\in \left\{ x_{k}|C_{1}\right\}
}N_{x_{k}}\right) CNOT\left( \left\{ I_{i}\right\} _{\in \left\{ all\text{ }%
x_{i}|C_{1}\right\} },I_{0}\right) \left( \prod_{x_{k}\in \left\{
x_{k}|C_{1}\right\} }N_{x_{k}}\right) N_{x_{0}},
\end{equation}
which requires one C$^{k}$-NOT gate and several NOT gates. Since it is
proved that the two-qubit C-NOT gate and single-qubit gates consist of a set
of univeral quantum logic gates, we use them to investigate the size of
these circuits. Iterating the construction of C$^{m}$-NOT from C$^{m-1}$-NOT%
\cite{Nielsenbook,Preskill}, the overall $7(m-1)$ gates including $3(m-1)$
C-NOT and $4(m-1)$ single-qubit gates are needed to obtain a C$^{m}$-NOT
gate. Accordingly, plusing at most $2m$ NOT gates, the overall circuit
complexity is $O(m)$. Similarly, the quantum circuit for a $k$-SAT problem
with one clause needs only $O(k)$ gates. In these two cases of SAT, $k$ and $%
m$ are of the same order $O(n)$ so that these QBCs can simulate such a SAT
problem with the polynomial computational cost in a single run.

According to the circuit configuration of Eqs. (3) and (4), the QBC for
random $k$-SAT problem with $m$ clauses ($m>1$) requires $m$ C$^{k}$-NOT,
one C$^{m}$-NOT and at most $km$ NOT gates, so that its size of $O(km)$ is
almost equal to the size of the $k$-SAT formula. For instance, a NP-complete
3-SAT needs $3(3m-2)$ C-NOT, $4(3m-1)$ single-bit and at most $3m$ NOT
gates. However, such a QBC has to involve $m$ additional qubits as the
scratchpads which would waste vast space and increase the difficulty of this
procedure. Economizing on space is also an important aspect to investigate
the efficiency of the circuit. The simplification by the logic rules\cite
{Lee,Iwama} might reduce the complexity of the QBC and optimize its space at
some extent. It is also possible to decrease the number of scratch-pad
qubits by the ``refreshed'' scheme\cite{Preskill,Ben73}. Concretely, we can
divide the computation into smaller steps of roughly equal size, run each
step, copy the output, then run this step backward to clear up the
scratch-pad qubits which can be used in the next steps due to the {\it %
reversibility} of quantum circuit\cite{Preskill,Ben73}. The requirement of
the step division is to remain the input unchanged before and after each
step, which can easily satisfied in our QBC construction. So the computation
can proceed to completion by constructing a `recursive' procedure to
generate the minimal amount of scratch-pad bits. Further, random SAT
problems are analogous to random $k$-SAT problems and their circuit sizes
are less than the corresponding $k_{\max }$-SAT where $k_{\max }$ is the
maximal number of variables in clauses.

Therefore, random SAT problems can be simulated by quantum circuits. Using
qubits instead of bits, we give a paralleization scheme for solving random
SAT problems through these QBCs with truly mixed states as the inputs other
than pseudo-pure states in a NMR quantum processor.

\subsection{Quantum Parallelization scheme for SAT Problems with truly mixed
states}

In NMR quantum computing, {\it n} spin-$\frac{1}{2}$ nuclei are chosen as
computational qubits $I_{1}$,$I_{2}$...,$I_{n}$ representing {\it n}
variables in the SAT problems, one as the work qubit $I_{0}$ and some
additional qubits as scratchpad qubits $I_{s_{\mu }}$ if needed. Just as the
general computation, our scheme is divided into three steps: (1) the
preparation of the initial state; (2) the process of computation; and (3)
the measurement of the computational results.

Step 1. We prepare the NMR ensemble into a mixed superposition state $\rho
_{in}$ of $2^{n}$ equally populated states $I_{n}^{\alpha /\beta
}...I_{2}^{\alpha /\beta }I_{1}^{\alpha /\beta }$ multiplied by $%
I_{0}^{\alpha }I_{s_{n}}^{\alpha }...I_{s_{1}}^{\alpha }I_{s_{m}}^{\alpha }$%
: 
\begin{eqnarray}
\rho _{in} &=&I_{0}^{\alpha }I_{s_{1}}^{\alpha }I_{s_{2}}^{\alpha
}...I_{s_{m}}^{\alpha }\equiv (I_{n}^{\alpha }...I_{2}^{\alpha
}I_{1}^{\alpha }+I_{n}^{\alpha }...I_{2}^{\alpha }I_{1}^{\beta
}+...+I_{n}^{\beta }...I_{2}^{\beta }I_{1}^{\beta })I_{0}^{\alpha
}I_{s_{m}}^{\alpha }...I_{s_{2}}^{\alpha }I_{s_{1}}^{\alpha }  \nonumber \\
&\equiv &\sum_{x=0...0}^{1...11}\left( \left| x\right\rangle \left\langle
x\right| \otimes \left| x_{0}=0\right\rangle \left\langle x_{0}=0\right|
\otimes \left| 0...00\right\rangle \left\langle 0...00\right| \right)
\end{eqnarray}
if the QBC for a given SAT problem involves {\it m} scratchpad qubits $%
I_{s_{\mu }}$. Here, we adopt the convenient representation of the $i$-th
spin polarization operators in Liouville spin space\cite{Ernst}: 
\begin{eqnarray}
I_{i}^{\alpha } &=&\left| 0\right\rangle \left\langle 0\right| =\frac{1}{2}%
\left( 1_{2}+I_{iz}\right)  \nonumber \\
I_{i}^{\beta } &=&\left| 1\right\rangle \left\langle 1\right| =\frac{1}{2}%
\left( 1_{2}-I_{iz}\right)
\end{eqnarray}
which represent the spin-up and spin-down states, respecitvely, where $1_{2}$
is a $2\times 2$ unit matrix and $I_{iz}$ is the z component of Pauli
operators of the $i$-th spin. Note that the work qubit $I_{0}$ and all
scratchpad qubits $I_{s_{\mu }}$ are in the $\left| 0\right\rangle $ state
and all computational qubits $I_{i}$ are random states includes all $2^{n}$
assignments of {\it n} variables, neither pure nor pseudo-pure states. When
no scratchpad qubits are involved in the circuit constructed, such as the
1-SAT and {\it k}-SAT with one clause, the ``one-qubit'' model of quantum
computation proposed by Knill et al. is obtained\cite{Knill}.

Step 2. The $U_{F-C-NOT}$ gate for the QBC constructed above is carried out
on such an input state $\rho _{in}$ of Eq. (8) which is transformed to the
output state $\rho _{out}$:

\begin{eqnarray}
\rho _{out} &=&U_{F-C-NOT}\rho _{in}U_{F-C-NOT}^{\dagger }  \nonumber \\
&=&U_{F-C-NOT}\left( \sum_{x=0...0}^{1...11}\left( \left| x\right\rangle
\left\langle x\right| \otimes \left| x_{0}=0\right\rangle \left\langle
x_{0}=0\right| \otimes \left| 0...00\right\rangle \left\langle 0...00\right|
\right) \right) U_{F-C-NOT}^{\dagger }.  \nonumber \\
&=&\sum_{x=0...0}^{1...11}\left( \left| x\right\rangle \left\langle x\right|
\otimes \left| x_{0}=F\left( x\right) \right\rangle \left\langle
x_{0}=F\left( x\right) \right| \otimes \left| C_{m}\left( x\right)
...C_{2}\left( x\right) C_{1}\left( x\right) \right\rangle \left\langle
C_{m}\left( x\right) ...C_{2}\left( x\right) C_{1}\left( x\right) \right|
\right)  \nonumber \\
&=&\sum_{x\in \left\{ x|F\left( x\right) =0\right\} }\left( \left|
x\right\rangle \left\langle x\right| \otimes \left| x_{0}=0\right\rangle
\left\langle x_{0}=0\right| \otimes \left| C_{m}...C_{2}C_{1}\neq
1...11\right\rangle \left\langle C_{m}...C_{2}C_{1}\neq 1...11\right| \right)
\nonumber \\
&&+\sum_{x\in \left\{ x|F\left( x\right) =1\right\} }\left( \left|
x\right\rangle \left\langle x\right| \otimes \left| x_{0}=1\right\rangle
\left\langle x_{0}=1\right| \otimes \left|
C_{m}...C_{2}C_{1}=1...11\right\rangle \left\langle
C_{m}...C_{2}C_{1}=1...11\right| \right) .
\end{eqnarray}
Becuase of the AND operation between clauses, the values of all clauses $%
C_{1},C_{2},...,C_{m}$ stored in the states of {\it m} scratchpad qubits $%
I_{s_{\mu }}$ have to be equal to 1 in order to satisfy the formula $F$. It
can be seen from Eq. (9) that, the solutions of the SAT formula $F$, i.e.,
those states $\left| x\right\rangle \in \left\{ \left| x\right\rangle \text{ 
}|\text{ }F\left( x\right) =1\right\} $, are labelled by the state of the
work qubit $I_{0}$ $\left| x_{0}\right\rangle =\left| 1\right\rangle $. To
see the proper performance of these QBCs, let me see a simple 1-SAT with the
formula $F=\overline{x_{1}}\wedge x_{2}\wedge x_{3}$, through $U_{F-C-NOT}$
consisting of a C$^{3}$-NOT gate and two NOT gates with no scratchpad
qubits, 
\begin{equation}
\rho _{out}=\sum_{x\neq 110}\left( \left| x\right\rangle \left\langle
x\right| \otimes \left| x_{0}=0\right\rangle \left\langle x_{0}=0\right|
\right) +\left| 110\right\rangle \left\langle 110\right| \otimes \left|
x_{0}=1\right\rangle \left\langle x_{0}=1\right| ,
\end{equation}
which demonstrates that the solution of the 1-SAT is $\left|
110\right\rangle $.

Step 3. The last step of computation is to extract efficiently the results
of computation stored in the work qubit $I_{0}$. Quantum strong measurement
prohibits such a successful process. However, in some special cases,
ensemble weak measurements ever considered as disadvantages in NMR quantum
computing can be advantageous. According to the spectral implementation of a
quantum computer\cite{Madi}, 2$^{n}$ logic states of {\it n} computational
qubits are assigned to individual spectral resonance lines of the work qubit 
$I_{0}$ when all resonance lines are distinguishable. Moreover, it can be
recoginized\cite{Ernst} from Eq. (9) that after a $\pi /2$ detection pulse
on spin $I_{0}$ the $I_{0}^{\alpha }$ ($\left| x_{0}=0\right\rangle $) state
gives a positive absorptive peak in a NMR spectrum with properly set
reciever phase settings, while the $I_{0}^{\beta }$ ($\left|
x_{0}=1\right\rangle $) state gives a negative peak. As a result, we label
the positive domain as the {\it FALSE} space and the negative {\it TRUE}
space. All scratchpad qubits $I_{s_{\mu }}$ should be decoupled during the
measurement because the solutions are irrelevant to these qubits. The result
is put into evidence by the sign of the resonance of spin $I_{0}$, i.e., the
solutions of a SAT problem corresponding to the positive resonance lines of
spin $I_{0}$ after a $\pi /2$ detection pulse should be found in the {\it %
TRUE} space if there exists the solution. It is, of course, necessary to
obtain a reference signal against which the phases of other signals of
interst can be determined, but this is easily achieved by acquiring a NMR
signal from thermal equilibrium $I_{0z}$ of spin $I_{0}$.

\section{Experiments}

The physical system used in our experiments is the carbon-13 labeled alanine 
$NH_{3}^{+}-C^{\alpha }H(C^{\beta }H_{2})-C^{^{\prime }}O_{2}^{-}$ dissolved
in $D_{2}O$. All experiments were performed on a Varian INOVA500
spectrometer with a probe tuned at 500.122MHz for $^{1}$H and at 125.768MHz
for $^{13}$C. The measured NMR parameters are listed in Table 1. Three $%
^{13} $C nuclear spins of 99\% abundance were chosen as a 3-qubit system
with decoupling all protons, while three $^{13}$C and one $^{1}$H nuclear
spins as a 4-qubit system with decoupling the methyl protons. In both of the
two separate systems, spin C$^{\alpha }$ was chosen as the work qubit $I_{0}$
due to its well-resolved scalar $J$ couplings to all other spins. Spins C$%
^{\prime }$, C$^{\beta }$ and $^{1}$H being directly joined with C$^{\alpha
} $, representing qubits $I_{1}$, $I_{2}$ and $I_{3}$. Their reference
spectra of thermal equilibrium are shown in Figure 3.

The key of sucessful experiments is how to implement effeciently the C$^{n}$%
-NOT gates which are elementary operations in these QBCs for SAT problems.
The C$^{n}$-NOT gate can be implemented in many ways. In all experiments,
the C-NOT gate was accomplished via the appropriate rotation and delay with
selective refocusing schemes\cite
{Price,Linden1,RefocusLin,RefocusLeu,RefocusJon}, while the C$^{n}$-NOT gate
when $n\geq 2$ by a single low-power transition-selective pulse\cite
{Ernst,Linden,Dorai,Jiang,Suter} which are very efficient and reliable on
our parallelization scheme becuase the input state remains always along z
axis when applying these transition-selective pulses\cite{PengBV}. In all
experiments, we employed a low-power {\it Gaussian} pulse of length 80$ms$
to selectively excite an individual transition. A technique was used to
realize a C$^{m}$-NOT gate in a more-qubit ($n>m+1$) system, i.e.,
decoupling the irrelevant qubits only when applying the line-selective
pulse. For example, to implement the C$^{2}$-NOT gate with spins C$^{\prime
} $ and C$^{\beta }$ as the control qubits and spin $I_{0}$ as the target
qubit in a 4-qubit system, we first decoupled spin $^{1}$H, then realized
the C$^{2}$-NOT gate just like that in a 3-qubit system and finally
recoupled spin $^{1}$H.

Begin with some simple speical cases of modified SAT problems such as $%
x_{1}\wedge \overline{x_{1}}$ and $x_{1}\vee \overline{x_{1}}$. As the AND
or OR\ operation connects the same variable or their negation, a scratchpad
qubit $I_{2}$ is introduced in the circuit to store their values by a C-NOT
gate. Then the scheme described in Sec. II was performed. First the mixed
state $I_{0}^{\alpha }I_{2}^{\alpha }$ was prepared from thermal equilibrium
by this following pulse sequence: $Y_{1}\left( \frac{\pi }{2}\right)
Y_{2}\left( \frac{\pi }{3}\right) -G_{z}-Y_{0}\left( \frac{\pi }{4}\right)
-\left( \frac{1}{8J_{02}}\right) -X_{1}\left( \pi \right) -\left( \frac{1}{%
8J_{02}}\right) -\overline{X_{1}}\left( \pi \right) X_{0}\left( \frac{\pi }{4%
}\right) ,$ where $Y_{i}(\theta )$ and $\overline{Y_{i}}(\theta )$ denote $%
\theta $ and $-\theta $ rotations about $\hat{y}$ axis on spin $i$ and so
forth, $\left( \frac{1}{8J_{02}}\right) $ describes a time evolution of $%
1/8J_{02}$ under the scalar coupling between spins $I_{0}$ and $I_{2}$ with
the selective refocusing $\pi $ pulses $X_{1}\left( \pi \right) $ and $%
\overline{X_{1}}\left( \pi \right) $\cite
{Price,Linden1,RefocusLin,RefocusLeu,RefocusJon} and $G_{z}$ represents a
pulsed-field gradient (PFG) along z-axis. The experimental results are shown
in Figure 4, (a) and (c) for the number of variables $n=1$ (on the left
column) on the 3-qubit homonuclear system; (b) and (d) for $n=2$ (on the
right column) on the 4-qubit system with qubit $I_{3}$ as the variable $%
x_{2} $. Due to the experimental limitation, we artificially subtract the
information related to spin $I_{2}$ by removing its relevant resonance
lines. There is no corresponding resonance lines in the {\it TRUE} space for
the case of $V_{1}\wedge \overline{V_{1}}$ (see figure 4 (a) and (b)),
indicating that no solutions exists in the problem of $x_{1}\wedge \overline{%
x_{1}\text{ }}$(i.e., $x_{1}\wedge \overline{x_{1}}\equiv 0$), but for $%
x_{1}\vee \overline{x_{1}}$ (see Figure 4(c) and (d)), all resonance lines
appear in the {\it TRUE} space, indicating that the logic expression $%
x_{1}\vee \overline{x_{1}}\equiv 1$. Furthermore, we also experimentally
tested other two cases: $x_{1}\wedge x_{1}\equiv x_{1}$ and $\overline{x_{1}}%
\wedge \overline{x_{1}}\equiv \overline{x_{1}}$. As shown in Figure 4
(e)-(h), the resonance lines related to the solutions are separated from the 
{\it FALSE} space, appearing in the {\it TRUE} space, which are in good
agreement with the theoretical expectations. In the same way, we can test
these simple formulas on a larger number of variables ($n>2$), i.e., to
search assignments satisfying these formulas in a larger database.

In addition, all the logic formulas of 1-SAT of two and three variables, and
the $k$-SAT with one clause were experimentally implemented on such a
3-qubit or 4-qubit NMR quantum processor. Qubits $I_{1}$, $I_{2}$ and $I_{3}$
are viewed as the variables $x_{1}$, $x_{2}$ and $x_{3}$, respectively. The
initial input state $I_{0}^{\alpha }$ from thermal equilibrium was more
easily prepared by applying radio-frequency (rf) $\pi /2$ pulses on all
other qubits except for $I_{0}$, followed by a PFG. Figure 5 shows the
spectra of spin $I_{0}$ after the execution of these quantum circuits for
all 1-SAT and 2-SAT of one clause with two variables. Figures 6 and 7 show
the experimental results of the partial problems for 1-SAT, 2-SAT and 3-SAT
of three variables. From these figures one can see that the solutions of a
given SAT problem are achieved from the resonance lines in the {\it TRUE}
space, as theoretically expected. Accordingly, the parallel computation for
SAT problems with mixed states succeeds in NMR ensemble quantum computing.

Besides, from these figures, we can also see the small but significant
distortions of the phases arising from the difficulty of implementing
perfect selective pulses and inhomogeneity of the static magnetic field. It
can be also seen that the distortions become larger as the system size is
increased. In addition, the signal decay is mainly due to the relaxation
effect of the low-power, long-duration transition-selective excitation.

\section{Conclusion}

In summary, we give a general construction of QBCs for random SAT problems
by using C$^{n}$-NOT and NOT gates and present a parallelization scheme for
solving them with mixed states on a NMR quantum information processor. This
model is an extension of ``one-qubit'' quantum computation. The present
model uses quantum parallelism to perform computation in an essential way
and then uses the topological nature of the NMR spectrum to obtain the
computational results by monitoring spin $I_{0}$ with a single detection
pulse, which in principle makes random NP-complete SAT problems be solved in
polynomial time. Meanwhile, we have experimentally implemented the scheme
for several typical SAT problems on a 3-qubit or 4-qubit NMR quantum
processor. The experimental results are well consistent with the theoretical
expectations. Compared to Hogg's algorithm\cite{Hogg,Zhu,Peng}, this present
scheme with a truly mixed input state avoids the pseudo-pure state
preparation in NMR quantum computing so as to simplify the process of the
whole experiment to solve such a problem. Moreover, its feasibility is not
limited in the restricted case of SAT (e.g., Hogg's algorithm for 1-SAT and
maximally constrained $k$-SAT), but for random SAT problems. However, it
will require some additional qubits as scratchpads, which may increase the
potential difficulty of finding a suitable sample. In principle, we can
build up a recursive (``refresh'') procedure to minimize the number of
scratchpads for saving space and the run-time cost. In order that all
database can be distinguishable, the chosen sample is required to have a
spin which has resolved scalar $J$ couplings to all other computational
spins. Though liquid-state NMR faces the difficulty of the scalability, the
scheme of effectively finding the solutions to random SAT problems with a
great many variables presented in this paper can not only be realized in NMR
quantum processors with a few qubits, but also might be feasible or
instructive for other scalable quantum-processor implementations with
similar features to the liquid-state NMR counterpart.

\begin{center}
{\bf ACKNOWLEDGMENTS}
\end{center}

We would like to thank Prof. D. Suter's suggestion for Hogg's algorithm with
mixed states. We also thank Hanzeng Yuan, Zhi Ren and Xiaowen Fang for help
in the course of experiments. This work is supported by the National
Fundamental Research Program (2001CB309300) and National Nature Sciences
Foundation of China (10274093).

Table 1. Measured NMR parameters for alanine dissolved in $D_{2}O$ on a
Varian INOVA500 spectrometer with respect to transmitter frequencies of
500.122MHz for $^{1}$H and 125.768MHz for $^{13}$C.

\bigskip

\begin{tabular}{cccccc}
\hline\hline
nuclei & $\nu /Hz$ & $J_{C^{^{\prime }}}/Hz$ & $J_{C^{^{\alpha }}}/Hz$ & $%
J_{C^{\beta }}/Hz$ & $J_{H}/Hz$ \\ \hline
$C^{\prime }$ & $-4320$ &  & $34.94$ & $-1.2$ & $5.5$ \\ \hline
$C^{\alpha }$ & $0$ & $34.94$ &  & $53.81$ & $143.21$ \\ \hline
$C^{\beta }$ & $15793$ & $-1.2$ & $53.81$ &  & $5.1$ \\ \hline
$H$ & $1550$ & $5.5$ & $143.21$ & $5.5$ &  \\ \hline\hline
\end{tabular}
\newpage

\begin{center}
{\large Figure Captions}
\end{center}

Figure 1.Quantun circuit of implementing the AND and OR gate by a Toffolli
gate. The top two bits ($x_{i},x_{j}$) represent the inputs, while the third
bit $x_{0}$ is initially prepared in the $\left| 0\right\rangle $ state as
the target bit to record the outputs. $\oplus $ denotes the modulo 2
addition (i.e., a NOT gate) and the controlling operation on a spin being in
the $\left| 1\right\rangle $ state is represented by a filled circle.

Figure 2. Quantum circuits for a 3-SAT problem with the formula being $%
\left( x_{1}\vee x_{2}\vee x_{3}\right) \wedge (x_{1}\vee x_{2}\vee 
\overline{x_{3}})\wedge \left( \overline{x_{1}}\vee x_{2}\vee x_{3}\right) $%
. Note two NOT gates in the dashed boxes can be canceled out pairwise. DEC
at the end of the circuit denotes that the qubit is decoupled in the signal
acquisition.

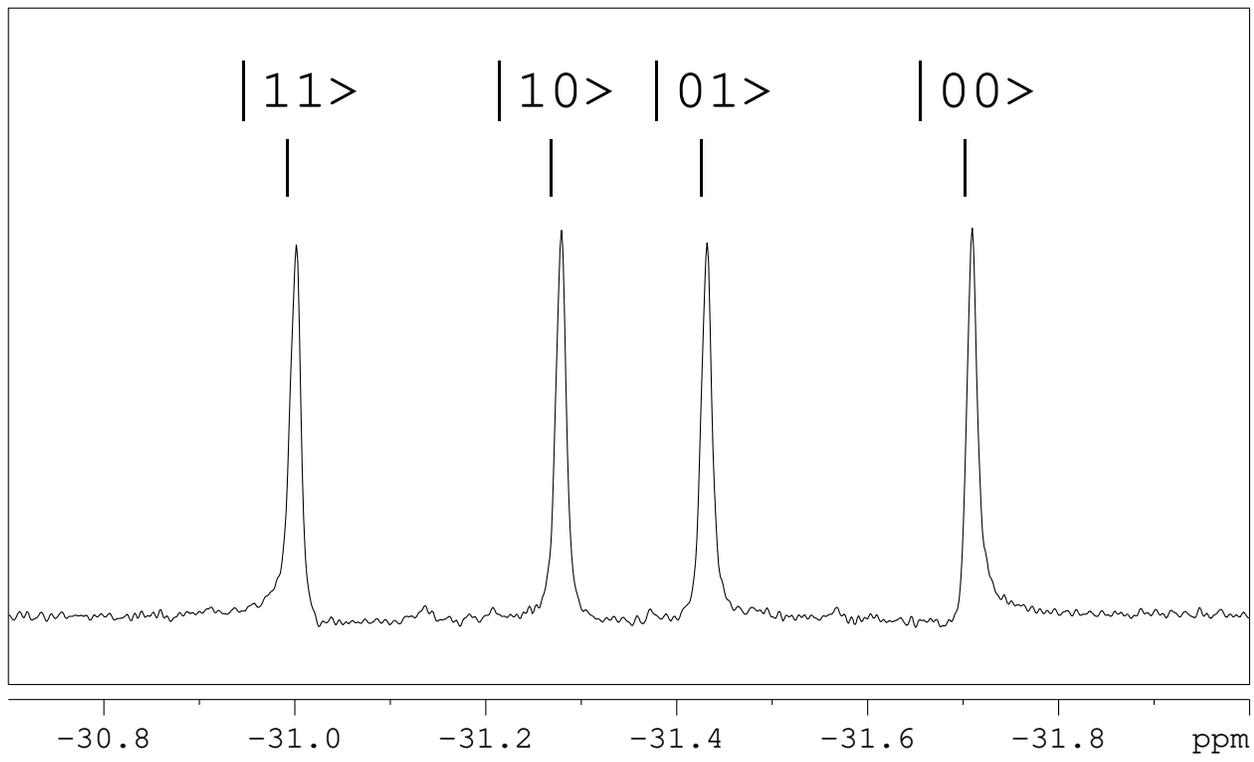
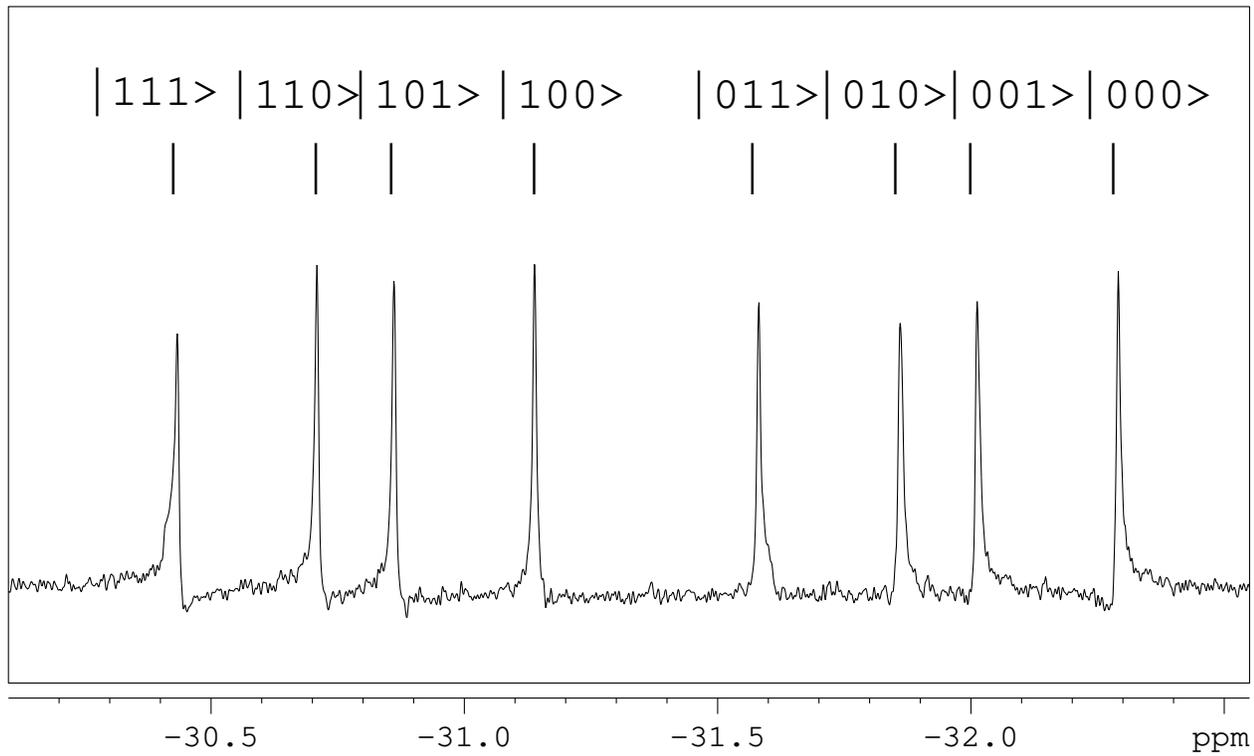
Figure 3. The reference spectra of spin $I_{0}$ in thermal equilibrium for
(a) a 3-qubit system and (b) a 4-qubit system. Split peaks are observed due
to spin-spin couplings. The labels of the peaks represent the states of the
other spins (see the text). All spectra in Figures 3-7 were recorded by a
single $\frac{\pi }{2}$ pulse on spin $I_{0}$.

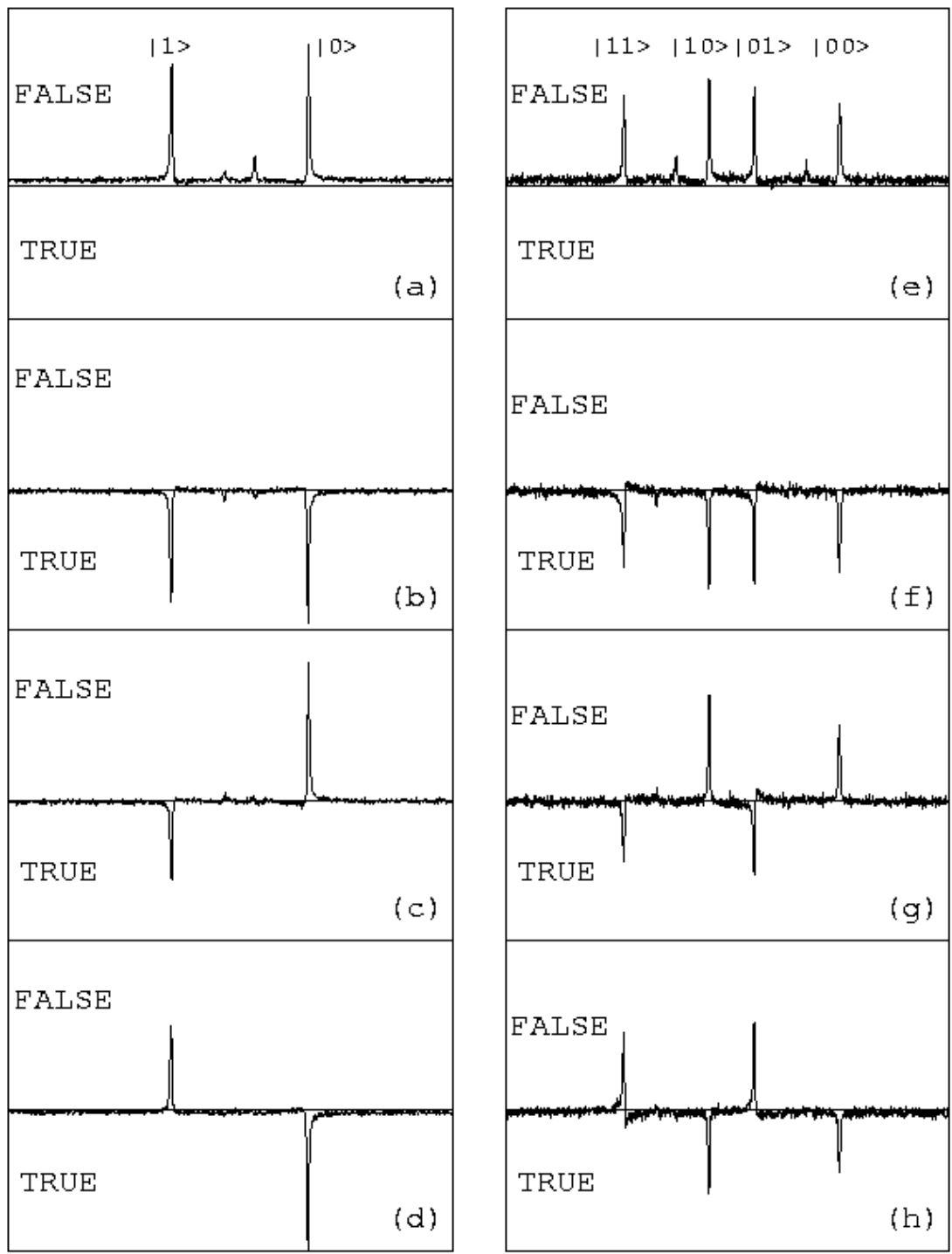
Figure 4. The experimental spectra of spin $I_{0}$ to investigate four
simple cases of modified SAT: $x_{1}\wedge \overline{x_{1}}$, $x_{1}\vee 
\overline{x_{1}}$, $x_{1}\wedge x_{1}$ and $\overline{x_{1}}\vee \overline{%
x_{1}}$ from the top to the bottom for $n=1$ (on the left column) and $n=2$
(on the right column).

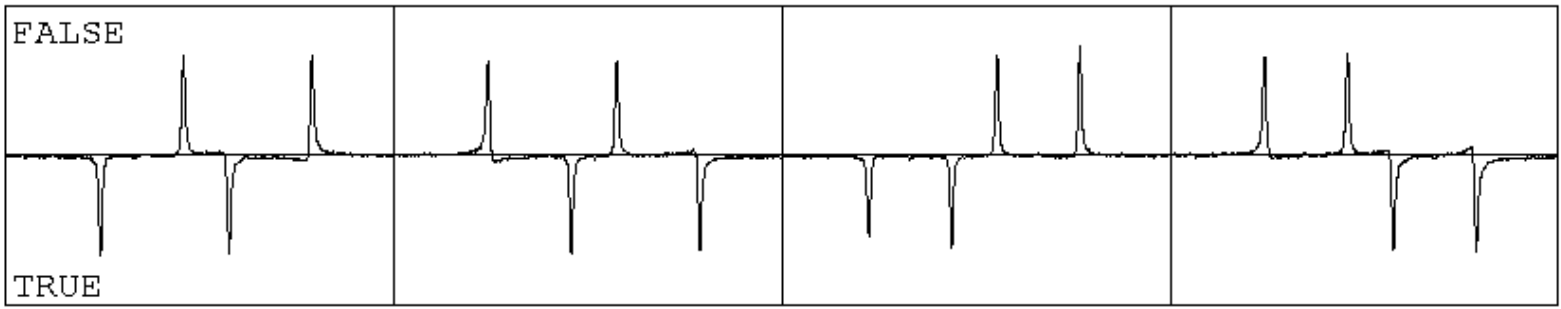
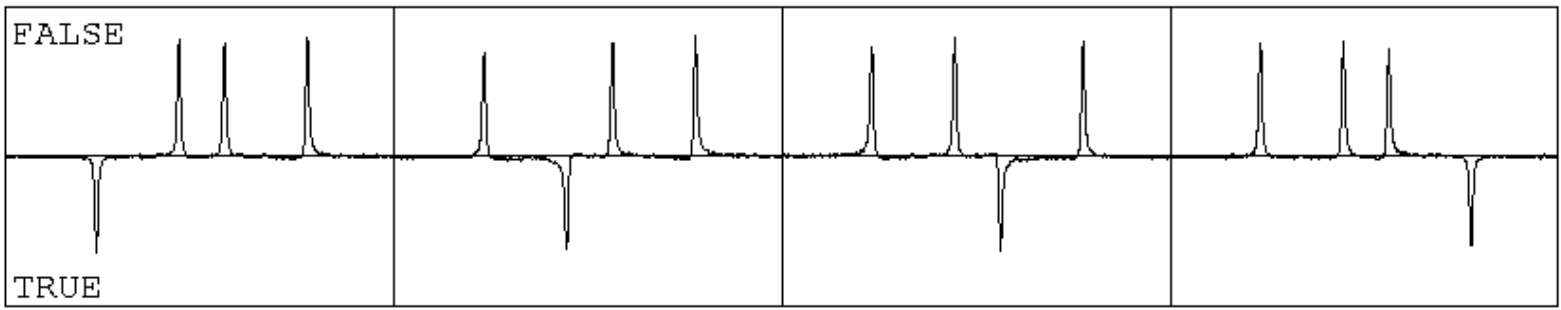
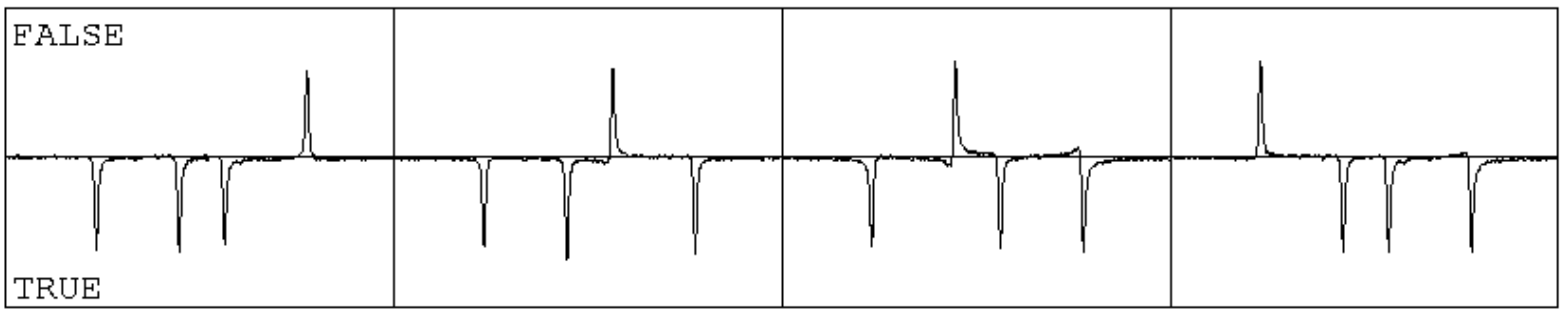
Figure 5. The experimental spectra of spin $I_{0}$ for all 1-SAT and 2-SAT
with one clause of two variables: (a) 1-SAT with $m=1$, the corresponding
logic formulas are $x_{1}$, $\overline{x_{1}\text{,}}$ $x_{2}\ $and $%
\overline{x_{2}}$; (b) 1-SAT with $m=2$, the corresponding the logic formula
are $x_{1}\wedge x_{2}$, $\overline{x_{1}}\wedge x_{2}$, $x_{1}\wedge 
\overline{x_{2}}\ $and $\overline{x_{1}}\wedge \overline{x_{2}}$; (c) 2-SAT
with $m=1$, the corresponding logic formulas are $x_{1}\vee x_{2}$, $%
\overline{x_{1}}\vee x_{2}$ $x_{1}\vee \overline{x_{2}}\ $and $\overline{%
x_{1}}\vee \overline{x_{2}}$; from left to right, respectively.

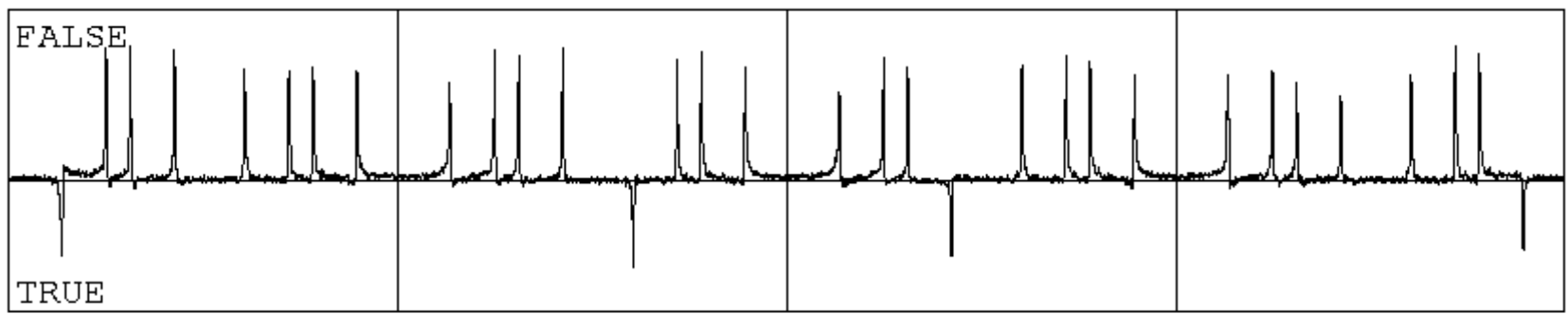
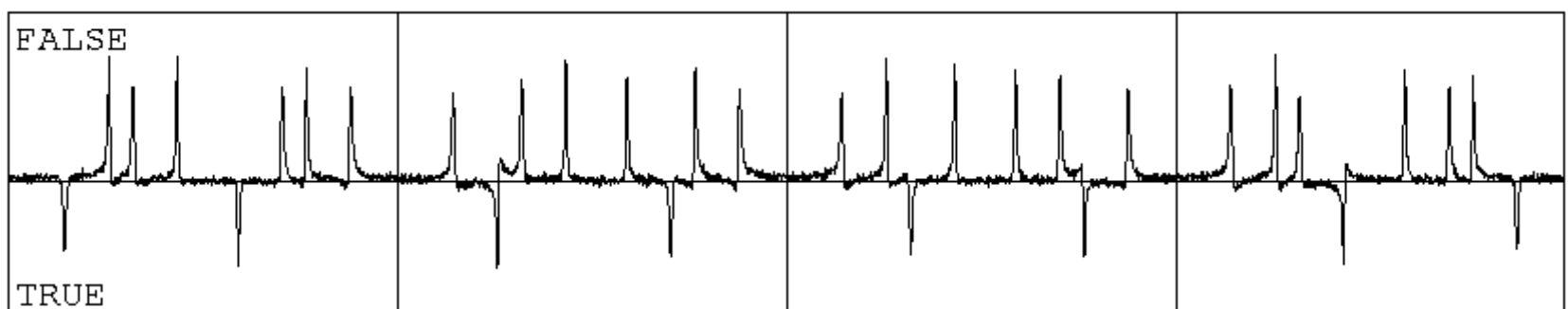
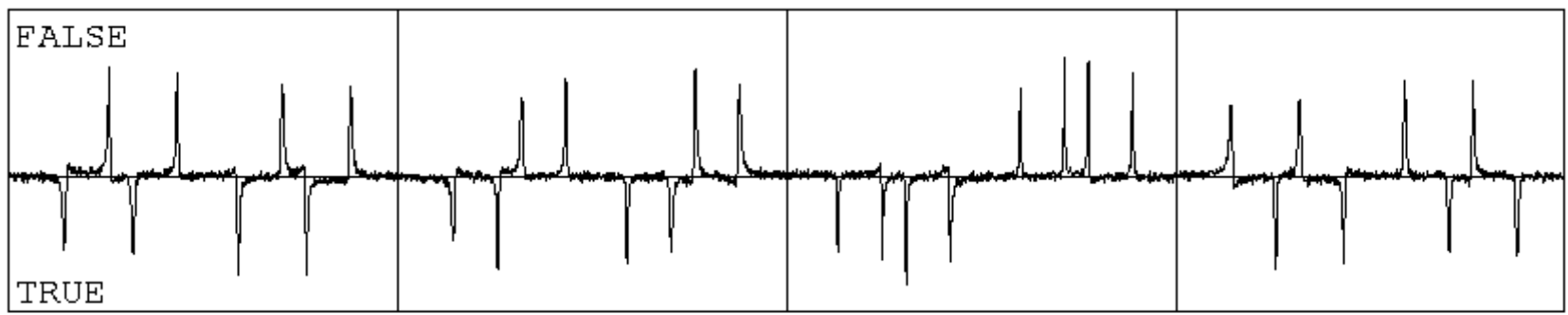
Figure 6. The experimental spectra of spin $I_{0}$ for the partial 1-SAT\
problems of three variables: (a) $m=3$, the corresponding logic formulas are 
$x_{1}\wedge x_{2}\wedge x_{3}$, $x_{1}\wedge x_{2}\wedge \overline{x_{3}}$, 
$\overline{x_{1}}\wedge \overline{x_{2}}\wedge x_{3}$ and $\overline{x_{1}}%
\wedge \overline{x_{2}}\wedge \overline{x_{3}}\ $; (b) $m=2$, the
corresponding logic formulas are $x_{1}\wedge x_{2}$, $\overline{x_{1}}%
\wedge x_{2}$ $x_{1}\wedge \overline{x_{2}}\ $and $\overline{x_{1}}\wedge 
\overline{x_{2}}$; (c) $m=1$, the corresponding logic formulas are $x_{1}$, $%
x_{2}$, $x_{3}\ $and $\overline{x_{1}}$; from left to right, respectively.

Figures 7. The experimental spectra of spin $I_{0}$ for all 3-SAT\ and the
partial 2-SAT of three variables with $m=1$: (a)-(h) 3-SAT, the
corresponding logic formulas are $x_{1}\vee x_{2}\vee x_{3}$, $\overline{%
x_{1}}\vee x_{2}\vee x_{3}$, $x_{1}\vee \overline{x_{2}}\vee x_{3}$, $%
\overline{x_{1}}\vee \overline{x_{2}}\vee x_{3}$, $x_{1}\vee x_{2}\vee 
\overline{x_{3}}$, $\overline{x_{1}}\vee x_{2}\vee \overline{x_{3}}$, $%
x_{1}\vee \overline{x_{2}}\vee \overline{x_{3}}$ and $\overline{x_{1}}\vee 
\overline{x_{2}}\vee \overline{x_{3}}$; (i)-(l) 2-SAT, the corresponding
logic formulas are $x_{1}\vee x_{2}$, $\overline{x_{1}}\vee x_{2}$, $%
x_{1}\vee \overline{x_{2}}$ and $\overline{x_{1}}\vee \overline{x_{2}}$,
respectively.$\ $


\begin{references}
\bibitem{Garey}  M. R. Garey and D. S. Johnson, Computers and
Intractability: a Guide to the Theory of NP-Completeness, Freeman, San
Francisco, (1979).

\bibitem{Cook}  S. Cook, in Proceedings of the 3rd Annual ACM Symposium on
Theory of Computing (ACM, New York, 1971), p. 151.

\bibitem{Deutsch}  D. Deutsh, and R. Jozsa, Proc. Roy. Soc. Lond. A, {\bf 439%
}, 553 (1992).

\bibitem{Cleve}  R. Cleve, A. Ekert, C. Macchiavello and M. Mosca, Proc.
Roy. Soc. Lond. A, {\bf 454}, 339 (1998).

\bibitem{Shor}  P.Shor, Algorithms for quantum computation: discrete
logarithms and factoring. proc. 35th Annu. Symp. on Found. of Computer
Science, (IEEE comp. Soc. Press, Los Alomitos, CA. 1994) 124-134.

\bibitem{Grover}  L. K. Grover, Phys. Rev. Lett. {\bf 79}, 325 (1997).

\bibitem{Hogg}  T. Hogg, Phys. Rev. Lett. {\bf 80,} 2473 (1998).

\bibitem{Zhu}  X. Zhu, X. Fang, M. Feng, F. Du, K. Gao, and X. Mao, Physica
D {\bf 156}, 179 (2001).

\bibitem{Peng}  X. Peng, X. Zhu, X. Fang, M. Feng, K. Gao and M. Liu, Phys.
Rev. A 65, 0423015 (2002).

\bibitem{Gershenfeld}  N. Gershenfeld and I. L. Chuang, Science {\bf 275},
350 (1997).

\bibitem{Cory1}  D. G. Cory, A. F. Fahmy and T. F. Havel, Proc. Natl. Acad,
Sci. USA {\bf 94,} 1634 (1997).

\bibitem{Knill}  E. Knill and R. Laflamme, Phys. Rev. Lett. 81, 5672 (1998).

\bibitem{Madi}  Z. L. Madi, R. Bruschweiler and R. R. Ernst, {\it J. Chem.
Phys.} 109, 10603 (1998).

\bibitem{Xiao}  L. Xiao and G. L. Long, Phys. Rev. A 66, 052320 (2002).

\bibitem{Nielsenbook}  M. A. Nielsen and I. L. Chuang, {\it Quantum
Computation and Quantum Information} (Cambridge Univ. Press, Cambridge,
2000).

\bibitem{Lee}  J.-S. Lee, Y. Chung, J. Kim, and S. Lee, Arxiv:
quant-ph/9911053.

\bibitem{Preskill}  J. Preskill, http://www.theroy.caltech.edu/\symbol{126}%
preskill/ph229.

\bibitem{Iwama}  Kazuo Iwama, Yahiko Kambayashi, and Shigeru Yamashita,
http://www.acm.org/sigs/sigda/Archives/ProceedingArchives/Dac/Dac2002/papers/2002/dac02/pdffiles/28\_4.pdfhttp://www.acm.org/sigs/sigda/Archives/ProceedingArchives/Dac/Dac2002/papers/2002/dac02/pdffiles/28\_4.pdfhttp://www.acm.org/sigs/sigda/Archives/ProceedingArchives/Dac/Dac2002/papers/2002/dac02/pdffiles/28\_4.pdf

\bibitem{Ben73}  C. H. Bennett, {\it IBM J. Res. Dev.} {\bf 17}, 525 (1973);
C. H. Bennett, {\it IBM J. Res. Dev.} {\bf 17}, 525 (1989).

\bibitem{Ernst}  R. Ernst, G. Bodenhausen and A. Wokaun, {\it Principles of
Nuclear Magnetic Resonance in One and Two Dimensions} (Oxford Univ. Press,
Oxford, 1990).

\bibitem{Price}  M. D. Price, S. S. Somaroo, A. E. Dunlop, T. F. Havel, and
D. G. Cory, Phys. Rev. A 60, 2777 (1999).

\bibitem{Linden1}  N. Linden, B. Herver, R. J. Carvajo and R. Freeman, Chem.
Phys. Lett. 311, 321 (1999).

\bibitem{RefocusLin}  N. Linden, B. Herver, R. J. Carvajo and R. Freeman,
Chem. Phys. Lett. 305, 28 (1999).

\bibitem{RefocusLeu}  D. W. Leung, I. L. Chuang, F. Yamaguchi and Y.
Yamamoto, Phys. Rev. A 61, 042310 (2002).

\bibitem{RefocusJon}  J. A. Jones and E. Knill, J. Magn. Reson. 141, 322
(1999).

\bibitem{Linden}  N. Linden, H. Barjat and R. Freeman, Chem. Phys. Lett. 
{\bf 296}, 61 (1998).

\bibitem{Dorai}  K. Dorai, Arvind and A. Kumar, Phys. Rev. A {\bf 61,}
042306 (2000).

\bibitem{Jiang}  J. Du, M. Shi, X. Zhou, Y. Fan, B. Ye, and R. Han, Phys.
Rev. A {\bf 64}, 042306 (2001).

\bibitem{Suter}  K. Dorai and D. Suter, quant-ph/0211030.

\bibitem{PengBV}  X. Peng, X. Zhu, X. Fang, M. Feng, K. Gao, and M. Liu,
ArXiv: quant-ph/0202008.\newpage
\end{references}
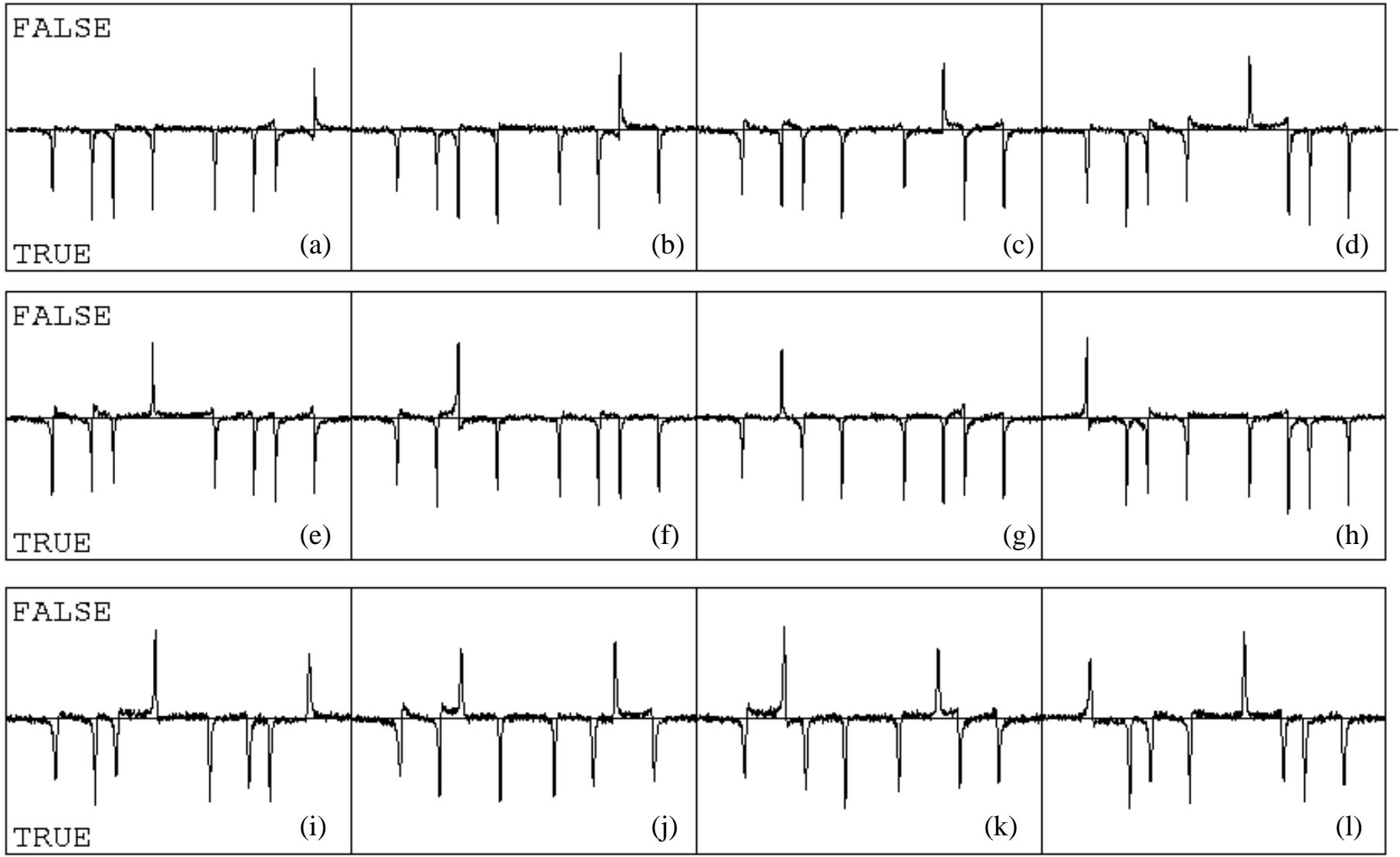
\end{document}